
\magnification=1200
\hsize 15.0 true cm
\vsize 23.0 true cm
\def\c{\centerline}
\def\v{\vskip 1pc}
\def\ej{\vfill\eject}
\def\b{{\bf b}}
\def\r{{\bf r}}
\def\p{{\bf p}}
\def\pperp{{{\bf p}_\perp}}
\def\z{{\hat{\bf z}}}
\def\En{{\cal E}}
\def\E{{\bf E}}
\def\B{{\bf B}}
\def\ra{\rangle}
\def\la{\langle}
\def\re{{\rm e}}
\def\half{{1 \over 2}}
\def\da{\dagger}
\overfullrule 0 pc
\
\vskip 1.5pc
\parindent 3pc\parskip 1pc
\c{\bf Back-reaction in a cylinder}
\vskip 4 pc
\c{J.M. Eisenberg$^*$}
\v
\c{\it School of Physics and Astronomy}
\c{\it Raymond and Beverly Sackler Faculty of Exact Sciences}
\c{\it Tel Aviv University, 69978 Tel Aviv, Israel}
\c{\it and}
\c{\it TRIUMF, 4004 Wesbrook Mall, Vancouver, B.C., Canada V6T 2A3}

\vskip 4pc
\noindent {\bf Abstract:-}  A system is studied in which initially a
strong classical electric field exists within an infinitely-long
cylinder and no charges are present.  Subsequently, within the cylinder,
pairs of charged particles tunnel out from the vacuum and the current
produced through their acceleration by the field acts back on the field,
setting up plasma oscillations.  This yields a rough model of phenomena
that may occur in the pre-equilibrium formation phase of a quark-gluon
plasma.  In an infinite volume, this back-reaction has been studied
in a field-theory description, and it has been found that the results of
a full calculation of this sort are well represented in a much simpler
transport formalism.  It is the purpose here to explore that comparison
for a situation involving a cylindrical volume of given radius.
\vfill
\noindent PACS numbers 11.15.Kc, 12.20.Ds
\vfill
\noindent October, 1994.
\vfill
\noindent $^*$Email: judah@giulio.tau.ac.il
\ej

\baselineskip 15 pt
\parskip 1 pc \parindent 3pc

\c{\bf I. INTRODUCTION}

The problem of back-reaction arsies in a number of contexts ranging from
the modeling of the pre-equilibrium phase of quark-gluon plasma
production [1,2] to studies of inflationary cosmology [3-5].  In general
terms, back-reaction consists in a situation where a system can be
viewed as governed by the mutual interaction between a field, usually
taken as classical, and pairs of charged particles that are produced
through its presence.  In one version, the initial condition posits the
pre-existence of the field and an approximate vacuum insofar as the
particles are concerned.  Pairs can then tunnel out of this vacuum by
reason of their interaction with the field in a purely quantal process.
These pairs are accelerated by the field, producing a current which in
turn acts back on the field.  The back-reaction leads to plasma
oscillations as one of its key characteristic effects.

Within the past few years, it has proved possible to carry out detailed
calculations for back-reaction in which boson [6] or fermion [7]
pairs are produced, including the case where boost-invariant coordinates
are used [8] as is appropriate for the study of the quark-gluon plasma.
(See also the didactic article in Ref. [9] and the review in Ref. [10].)
Somewhat surprisingly, it has emerged that calculations
[1,2] based on the much simpler transport formalism follow the
field-theory results in detail provided that quantum fluctuations are
smoothed and quantum statistical effects are introduced into the
transport equations [6-10].  That is, it proves possible to mimic the
quantum tunneling through the use of a source term based on the
Schwinger mechanism [11] in a transport equation.  The squares of the
quantal mode amplitudes, describing the momentum distribution of the
produced pairs, even agree remarkably closely with the corresponding
transport-function momentum dependence once smoothing and statistics are
included.  There are also formal derivations [12,13] of the mapping from
field theory to transport formalism in this context, providing [13], for
example, an understanding of how nonmarkovian features enter the theory.
Thus the back-reaction problem provides a model of a system sufficiently
simple that the nonperturbative field-theory description can be fully
solved numerically and approximations to it---notably the transport
formalism---can be studied in detail, but with enough complexity to
allow for a richness of phenomena and features.  This is
especially interesting when the nonequilibrium thermodynamics of
back-reaction is considered [8].  The field-theory version then refers
to the time evolution of a pure quantum state, while the transport
formalism has a well-defined entropy which increases with time.

One may also hope that essential features of the quark-gluon plasma
lacking in this simple back-reaction model will eventually be
incorporated and their effects studied.  These must surely include (i)
the nonabelian nature of quantum chromodynamics which governs the
plasma; (ii) the nonclassical nature of the chromoelectric field that
enters there; and (iii) the fact that the plasma lives in a finite
volume, not in the infinite volume treated so far in back-reaction.  To
the degree that a mapping from field theory to transport formalism
can be found,
say, in the presence of each of these features separately, the use of
the relatively simple transport formalism for the description of the
plasma gains considerable support.  Towards this end, some progress has
been made in the formulation of back-reaction for the nonabelian
Yang-Mills case [14]; a detailed study [15] has been carried out for the
lowest quantal correction to the classical electric field used so far;
and it is the purpose of this paper to study back-reaction in a
(transversely) finite volume of a geometry chosen for its eventual
applicability to the quark-gluon plasma.

We consider back-reaction in an infinitely-long cylindrical sleeve, as
shown in Figure 1, for an abelian and classical electromagnetic field.
The choice of a volume whose finiteness is only in the direction
transverse to that of the electric field and particle current was made
with a view to eventual extension to the quark-gluon plasma, where
boost-invariant coordinates would be used for the longitudinal
variables.  (In one spatial dimension, the case of back-reaction on a
finite line segment has been considered [16] in the past.)  The symmetry
here about the cylindrical axis is, of course, a sizable simplification
for our problem, though it still leaves in it the complexities of mode
mixing and of renormalization, and appears to be just barely amenable to
handling at the numerical level with present 100-MHz or 150-MHz computer
processors.  We develop below the formalism for back-reaction in this
geometry, first for field theory and then for the transport formalism,
then discuss briefly numerical methods and approximations for each, and
compare results for the two approaches.
\ej
\c{\bf II.  FIELD-THEORY FORMALISM}

We consider QED for scalar particles of charge $e$ and mass
$m,$ governed by the Klein-Gordon equation
$$[(\partial^\mu + ieA^\mu)(\partial_\mu + ieA_\mu) + m^2]\Phi = 0,
\eqno(1)$$
coupled with the classical Maxwell equation
$$\bigg({\partial^2 \over \partial t^2} - \nabla^2\bigg)A^\mu
= \la j^\mu \ra.\eqno(2)$$
Here $A^\mu$ is the classical four-vector potential and $j^\mu$ is the
four-current, taken in an expectation value for some initial
configuration of the charges.  The current is given by
$$\eqalign{j^\mu & = \half ie\{\Phi^\da (\partial^\mu + ieA^\mu)\Phi
- [(\partial^\mu + ieA^\mu)\Phi]^\da \Phi \cr
& - \Phi[(\partial^\mu + ieA^\mu)\Phi]^\da
+ [(\partial^\mu + ieA^\mu)\Phi] \Phi^\da\},}\eqno(3)$$
in terms of a second-quantized field $\Phi.$

We choose the geometry of an infinite cylinder of radius $R$ (see Figure
1) and restrict the initial conditions to be independent of the polar
angle $\varphi$ and assume that initially there are no regions of
nonvanishing net charge $j^0(\r,t)$ and that the initial field
configuration consists of $\E(b,t)\parallel \z,$ where $b=|\b|$ is the
radial coordinate.  The current ${\bf j(\r,t)}$ then remains parallel to
the axis of the cylinder $\z.$  This is because (see Figure 1) the
opposite charges of a produced pair are accelerated by $\E$ in opposite
directions parallel to $\z.$  The magnetic field $\B(\b,t)$ that is
produced has solenoidal geometry, whence both the positive and the
negative charges undergo transverse acceleration of the same direction
and magnitude, canceling any transverse contribution to the current.
Selecting a gauge such that $A^\mu = (0;0,0,A(b,t))$ has a $z$-component
only, the Klein-Gordon equation becomes
$$\bigg[{\partial^2 \over \partial t^2} - \nabla_\perp^2
-\bigg({\partial \over \partial z} - ieA(b,t)\bigg)^2 + m^2\bigg]
\Phi(b,\varphi,z;t) = 0,\eqno(4)$$
where
$$\nabla_\perp^2 = {\partial^2 \over \partial b^2}
+ {1 \over b}{\partial \over \partial b} + {1 \over b^2}
 {\partial^2 \over \partial \varphi^2}.\eqno(5)$$

We choose boundary conditions such that there are no particles
on or outside the cylinder wall at $b=R$; the electric field is also
taken to vanish at and beyond $b=R.$  These are boundary conditions
appropriate to electromagnetism.  One could of course choose bag
boundary conditions in their place.  In
expanding $\Phi$ over creation and annihilation operators, we then
use the orthonormal basis set
$$\phi_{ln}(\b) = {\re^{il\varphi} \over \sqrt{2\pi}}{\sqrt{2} \over R}
 {J_l(k_\perp(ln)b) \over J'_l(k_\perp(ln)R)},\eqno(6)$$
where $k_\perp(ln) = z_{ln}/R,$ $z_{ln}$ being the $n$th zero of the
Bessel function of order $l.$  In the following, we indicate the pair
$\{l,n\}$ by $n$; the lack of $\varphi$-dependence restricts our interest
to $l = 0$ in any event.  Then
$$\Phi = \sum_n \int_{-\infty}^\infty {dk \over 2\pi} \phi_n(b)
 {\re^{ikz} \over (2\omega_{k,n}^0)^{1/2}}
\big[a_{k,n}(t) + b_{-k,n}^\da(t)\big],
\eqno(7)$$
where $a_{k,n}$ annihilates particles of longitudinal momentum $k$ and
radial mode $n,$ $b_{-k,n}^\da$ creates antiparticles with $-k$ and
$n,$ and $\omega_{k,n}^0 = [k^2 + k_\perp^2(n) + m^2]^{1/2}.$  The usual
commutators pertain,
$$[a_{k,n}(t),\, a^\da_{k',n'}(t)] = [b_{k,n}(t),\, b^\da_{k',n'}(t)] =
2\pi \delta(k - k') \delta_{n,n'},\eqno(8{\rm a})$$
$$[a_{k,n}(t),\, b^\da_{k',n'}(t)] = 0,\eqno(8{\rm b})$$
and so forth.  We must relate these operators to their values at $t = 0,$
which we do by means of a Bogolyubov transformation [17]
$$a_{k,n}(t) = \sum_{n'}[u_{k,nn'}(t) a_{k,n'}(0)
+ v_{k,nn'}(t) b_{-k,n'}^\da(0)],\eqno(9{\rm a})$$
and
$$b_{-k,n}^\da(t) = \sum_{n'}[v_{k,nn'}^*(t) a_{k,n'}(0)
+ u_{k,nn'}^*(t) b_{-k,n'}^\da(0)],\eqno(9{\rm b})$$
where
$$u_{k,nn'}(0) = \delta_{nn'},\quad\quad v_{k,nn'}(0) = 0,\eqno(9{\rm c})$$
$$\sum_{\bar n}[u_{k,n\bar n}(t) u_{k,n'\bar n}^*(t)
- v_{k,n\bar n}(t) v_{k,n'\bar n}^*(t)] = \delta_{nn'},\eqno(9{\rm d})$$
and
$$\sum_{\bar n}[u_{k,n\bar n}(t) v_{k,n'\bar n}(t)
- v_{k,n\bar n}(t) u_{k,n'\bar n}(t)] = 0,\eqno(9{\rm e})$$
in an obvious extension of the usual [17] forms.

Defining the mode amplitudes
$$f_{k,nn'}(t) = {1 \over (2 \omega_{k,n}^0)^{1/2}}
[u_{k,nn'}(t) + v_{k,nn'}^*(t)],\eqno(10{\rm a})$$
with the initial values
$$f_{k,nn'}(0) = {1 \over (2 \omega_{k,n}^0)^{1/2}}\delta_{nn'},\quad
\dot f_{k,nn'}(0) = -i \bigg({\omega_{k,n}^0 \over 2}\bigg)^{1/2}
\delta_{nn'},\eqno(10{\rm b})$$
the scalar field becomes
$$\Phi = \sum_{nn'} \int_{-\infty}^\infty {dk \over 2\pi} \re^{ikz}
\phi_n(b)[f_{k,nn'}(t) a_{k,n'}(0) + f_{k,nn'}^*(t) b_{-k,n'}^\da(0)].
\eqno(11)$$
The normalization of the mode amplitudes in Eqs. (10) is taken so as
easily to accommodate the canonical commutation relations
$$[\Phi(\r,t),\ \Pi(\r',t)] = i\delta(\r-\r'),\eqno(12)$$
where $\Pi = \dot\Phi$ is the field conjugate to $\Phi.$  Equation (12),
with the commutators of Eqs. (8), leads, among other similar relations,
to the wronskian condition for the mode amplitudes,
$$\sum_{\bar n}[f_{k,n\bar n}(t) \dot f_{k,n'\bar n}^*(t)
- f_{k,n\bar n}^*(t) \dot f_{k,n'\bar n}(t)] = i\delta_{nn'},\eqno(13)$$
which is guaranteed by the dynamics to be satisfied at all times if it
is fulfilled initially, as it is of course by Eq. (10b).

Substituting Eq. (11) into Eq. (4) yields the dynamical equations for
the mode amplitudes
$$\ddot f_{k,nn'}(t) + \sum_{\bar n}\la n|\omega_k^2(t)|\bar n\ra
f_{k,\bar nn'}(t) = 0,\eqno(14{\rm a})$$
where
$$\eqalign{\la n|\omega_k^2(t)|\bar n\ra & =
\int d\b \ \phi_n^* \omega_k^2(\b,t)\phi_{\bar n} \cr
& = \int d\b \ \phi_n^*
\{\big(k - eA(b,t)\big)^2 + k_\perp^2(n)
+ m^2\}\phi_{\bar n}.}\eqno(14{\rm b})$$

For our geometrical conditions and with the assumption that the initial
charge configuration in Eq. (2) is the adiabatic vacuum, annihilated by
$a_{k,n}(0)$ and by $b_{k,n}(0),$ the expectation value of the current
becomes
$$\eqalign{\la j \ra = e \int & {dk \over 2\pi} \big(k - eA(b,t)\big)
\sum_{nn'} \phi_n(b) \phi_{n'}^*(b) \cr
& \times \sum_{\bar n}[f_{k,n'\bar n}^*(t) f_{k,n\bar n}(t)
+ f_{k,n'\bar n}(t) f_{k,n\bar n}^*(t)].}\eqno(15)$$

As is well known (see, e.g., the broad discussion in Ref. [17]), the
current in Eq. (15) suffers from divergences.  For the case of infinite
volume, this has been dealt with systematically through the procedure of
adiabatic regularization [17].  Unfortunately, that is inapplicable here
because of the mode mixing inherent in our problem and explicit in Eqs.
(14).  We therefore follow an alternative strategy here.  First we
consider the high-momentum limit ($k \to \infty,\ n \to \infty$) for
$f_{k,nn'}(t)$ satisfying the dynamical equations (14) and the
wronskian condition (13).  A direct generalization of the usual WKB
approach [17] suggests
$$f_{k,n'n}(t) \longrightarrow \la n'|{\exp[-i\int^t\omega_k(t')dt']
\over \sqrt{2\omega_k(t)}}|n\ra,\eqno(16)$$
as $k,n,n' \to \infty,$ where we restrict ourselves to the leading term
of the WKB since the higher-order refinements are of no use in the
presence of mode mixing.  It is easily seen that this form satisfies Eq.
(13) exactly and Eqs. (14) through order $k^0.$  To study the worst
divergence in Eq. (15), we substitute this into the factor involving a
summation over $n,$ $n',$ and $\bar n$ there to obtain, using closure,
$$\eqalign{\sum_{nn'} & \phi_n(b) \phi_{n'}^*(b)\cdot 2\pi\int_0^\infty
b' db' \phi_n^*(b') {1 \over \omega_k(b',t)}\phi_{n'}(b') = \cr
& = \sum_n \phi_n(b) \phi_n^*(b) {1 \over \omega_k(b,t)},}\eqno(17)$$
where we again used closure to obtain the last line.  It is then easy to
see that this divergent form in fact vanishes upon symmetric integration
[10,17] in Eq. (15).  This identifies the worst, cubic divergence there,
which is dealt with by subtracting the form of (17) to arrive at
$$\eqalign{\la j \ra & = e \int_{-\infty}^\infty {dk \over 2 \pi}
\big(k - eA(b,t)\big) \sum_{nn'} \phi_n(b) \phi_{n'}^*(b) \cr
& \times \bigg\{\sum_{\bar n}[f_{k,n'\bar n}^* f_{k,n\bar n}
+ f_{k,n'\bar n} f_{k,n\bar n}^*]
- {\delta_{nn'} \over \omega_k(b,t)}\bigg\}.}\eqno(18)$$

This expression still contains, of course, the usual logarithmic
divergence of charge renormalization, dealt with in adiabatic
regularization [10,17] by considering higher-order WKB than shown in Eq.
(16).  This is what does not go through in the mode-mixed case.
Instead, we base ourselves on the expectation that charge
renormalization is what must emerge in any event.  This is shown
explicitly for adiabatic regularization in the infinite volume in Ref.
[17], and we do not expect a change in high-momentum, short-interval
behavior here merely because of the presence of the cylinder walls.
Thus we calculate Eq. (18) with a $k$-momentum cutoff $\Lambda$
and a radial mode cutoff $n_{max},$
at the same time replacing the charge $e$ by the renormalized charge
$$\sqrt{Z}e = e (1 + e^2\ \delta e^2)^{-1/2} = e \bigg[1
+ {e^2 \over 24\pi^2}\log\bigg({1 \over m}
(\Lambda^2 + \pi^2 n_{max}^2/R^2)^{1/2}\bigg)\bigg]^{-1/2},\eqno(19)$$
and testing that results are insensitive to $\Lambda$ and $n_{max}.$

Equation (18) is now used in Eq. (2) as
$$\bigg({\partial^2 \over \partial t^2} - \nabla_\perp^2 \bigg)A(b,t)
= \la j(b,t) \ra\eqno(20)$$
for the dynamics of the electromagnetic field.  The coupled equations
that must be solved for the field-theory study of back-reaction in an
infinitely-long cylinder are then Eqs. (14), (18), and (20).
\v\v
\c{\bf III. TRANSPORT FORMALISM}

The transport equation we must deal with in the presence of the
cylindrical symmetry of Figure 1 is, for a positive charge $e,$
$$\eqalign{{\partial \over \partial t} f(\b;\p;t) &
+ {\pperp \over \En}\cdot{\partial \over \partial \b} f(\b;\p;t) \cr
& + e \bigg(\E + {\p \over \En}\times\B\bigg)\cdot
\bigg({\partial \over \partial \pperp}
+ \z {\partial \over \partial p_\parallel}\bigg) f(\b;\p;t)
= S\{E,B;\p\},}\eqno(21)$$
where $f$ is the transport function and $S$ is a source function based
on the Schwinger mechanism for pair production in a fixed field.  The
spatial variable $\b$ is, as hitherto, the transverse position measured
from the cylinder axis; again there is no $z$-dependence here.  The
momentum $\p = \{\pperp,\ p_\parallel\}$ is taken in terms of a transverse
component $\pperp$ and a longitudinal one $p_\parallel,$ while $\En =
(\pperp^2 + p_\parallel^2 + m^2)^{1/2}$ is the particle energy.  Note that
with our geometry there can be dependence only on an azimuthal angle
$\varphi,$ the angle between $\b$ and $\pperp.$  In parallel to the field
case of the previous section, we impose the boundary condition
$f(b=R;\ \pperp,p_\parallel;\ t) = 0.$  Given the field configuration
discussed in the previous section and shown in Figure 1, the transport
equation (21) takes on the form
$${\partial f \over \partial t}
+ {\pperp \over \En} \cdot {\partial f \over \partial \b}
+ e\bigg[E{\partial \over \partial p_\parallel}
+ \bigg({\pperp \over \En}\times\B\bigg)\cdot\z\
 {\partial \over p_\parallel}
+ \bigg({p_\parallel \over \En}\z \times \B\bigg)\cdot
 {\partial \over \partial \pperp}\bigg]f = S.\eqno(22)$$
Negative charges move in accordance with Eq. (21) or (22), but with $e$
replaced by $-e.$  Since the source term depends only on $|e|,$ it
follows from Eq. (22) that the transport function for negative charges
is given by that for positive charges with $p_\parallel\to -p_\parallel.$

The source term in Eq. (21) or (22) is taken as [6-10]
$$S\{E,B;\p\} = \delta(p_\parallel) |e| \sqrt{E^2 - B^2}
\log\bigg[1 + \exp\bigg(-{\pi (\pperp^2+m^2) \over |e|\sqrt{E^2 - B^2}}
\bigg)\bigg],\eqno(23)$$
where we insist that the invariant $E^2 - B^2$ be timelike, so that
energy considerations allow pairs to be
produced, taking $S = 0$ for $B \ge E.$  In Eq. (23), the usual
choice [6-10,12] has been made to the effect that the particles are
produced at zero longitudinal momentum.  A study based on the projection
method [13] has found that the source term is well localized around
$p_\parallel \sim 0$ for rather large fields, $e\sqrt{E^2 -
B^2}/\sqrt{\pperp^2+m^2} > 1$ in our notation; this situation is, in
fact, the only one considered here.  Since our results depend
ultimately on the pair current, which involves an integration over
$p_\parallel,$ they prove to be very insensitive to the appearance of
$\delta(p_\parallel)$ in $S,$ being essentially unchanged for
distributions in $p_\parallel$ centered around $p_\parallel = 0,$
normalized with respect to integration over $p_\parallel$ as is
$\delta(p_\parallel),$ and with widths ranging up to several times $m.$
The nonmarkovian features found in Ref. [13] also disappear, of course,
with the source term of Eq. (23), in consonance with the findings of
Ref. [13] for large fields.  Last, we shall not consider here the
consequences of Bose enhancement in the transport equation treatment,
the general effects being well known from previous work [6,10]; to
deal with such enhancement for the finite-volume case would require
some selection of cell size in phase space within which statistical
effects would take place.

The nontrivial Maxwell equations in our geometry are
$${\partial B \over \partial t} = {\partial E \over \partial b}\quad
 {\rm and}\quad
 {\partial E \over \partial t} = -j + {1 \over b}
 {\partial \over \partial b}(bB),\eqno(24{\rm a,b})$$
where $j$ is the current---again purely longitudinal---given by a
conduction and a polarization part,
$$j(b,t) = j_{\rm cond}(b,t) + j_{\rm pol}(b,t),\eqno(25{\rm a})$$
where
$$j_{\rm cond}(b,t) = 2e\int {d\p \over (2\pi)^3}
 {p_\parallel \over \En} f(\b,\p;t)\eqno(25{\rm b})$$
and
$$j_{\rm pol}(b,t) = {2 \over E} \int {d\p \over (2\pi)^3} \En
S\{E,B;\p\}.\eqno(25{\rm c})$$
The coupled equations that must be solved for the transport-theory study
of back-reaction are then Eqs. (22), (24), and (25).
\v\v
\c{\bf IV. NUMERICAL PROCEDURES}

We discuss first the procedures for the field-theory formulation.
Since the three-dimensional studies of back-reaction in an infinite
volume proved taxing from the computational point of view [10], it was,
of course, to be expected that the finite-volume case would present
difficulties in terms of the length of processor time required, and this
was indeed the case.  The infinite-volume calculation required very
refined grids for the time variable (typically the time step was on the
order of $10^{-4}\ m^{-1}$) and longitudinal momentum (typically
requiring some $10^3$ points).  The renormalization scheme used here
[see Eq. (19) and the discussion surrounding it] is, however, simpler
than the iterative one used for infinite volume [10].

Numerical procedures were straightforward: the current $j(b,t)$ and
potential $A(b,t)$ and field $E(b,t)$ were all expanded in terms of
the basis set of Eq. (6).  The initial field was usually taken in the form
$$E(b,0) = E_0 J_0(z_{00}b/R),\eqno(26)$$
in the notation of Section II.  The mode amplitudes $f_{k,nn'}(t)$ were
then calculated from the coupled ordinary differential equations (14)
using Runge-Kutta and the Fourier-Bessel components of the current
$j(b,t)$ were evaluated by Simpson integration of Eq. (18).  The
subtraction of the counter term in (18) required a return to
configuration space for the evaluation of $\omega_k^{-1}(b,t).$
The Fourier-Bessel components of the potential and field were
then advanced using Eq. (20)---again an ordinary differential equation
in the transform space---with Runge-Kutta.  Thus worries about possible
instabilities in partial differential equations were avoided.

The rapid fluctuations encountered in the mode amplitudes in the
infinite-volume case seem here to translate into a great deal of mode
coupling in the finite volume, with a consequent need for a large number
of Fourier-Bessel components in order to represent the current
adequately.  We here worked typically with about 500 longitudinal
momentum points and
summed $n$ out to about $n_{max} \sim 20.$  Time steps again had to be
kept small ($dt \sim 10^{-4}$).  Since computation times rise
quartically, or faster,
with $n_{max},$ this led to runs of several weeks with a 100-MHz or
150-MHz processor.  Moreover, as for infinite volume, quantum
fluctuations will eventually be more rapid than can be accommodated by
any grid choice, so that our computation scheme must eventually break
down.  Confidence in the validity of the field calculation is therefore
partly based on the usual tests checking that results do not change
appreciably when grid sizes are refined.  It also partly derives from the
qualitative and even semiquantitative agreement between field and
transport methods over the time range we treat, a comparison which is,
of course, the primary goal of this study.

Somewhat surprisingly, the transport formalism is also somewhat tedious
computationally.  The infinite-volume case allowed for solution of the
transport equations using essentially an analytic scheme based on the
method of characteristics [10], but for a finite volume we had to solve
Eq. (22) as a partial differential equation, using a Lax method [18].
Other methods, such as Lax-Wendroff or staggered leapfrog seemed to
offer little improvement over this.  The time evolution of the
electromagnetic fields as partial differential equations (24)---even in
the absence of a current $j$---proved unstable and this was not easily
cured through the use of Lax, Lax-Wendroff, or leapfrog methods, so we
again resorted to Fourier-Bessel decomposition and advanced the
coefficients in time using a predictor-corrector method.  The time
increment was again $dt \sim 10^{-4}$ and roughly 120 longitudinal
momentum points were required.  In the transverse direction about 120
momentum points and 80 spatial points were needed.

The outcome of all this was again very long computation time.  The
situation was eased somewhat by great insensitivity to the treatment of
the angular variable $\varphi.$  Eventually we chose to average over
this, fixing $\varphi = \pi/2.$  This changed the results typically by
only a few percent.  We also eventually averaged the transverse momentum
$|\pperp|$ in a manner suggested by the gaussian dependence of the source
term of Eq. (23), taking
$\la p_\perp \ra = (\sqrt{\pi |e|}/2)(E^2-B^2)^{1/4}.$
This approximation reproduced very well
the amplitudes and frequencies of the oscillatory behavior of $j(b,t)$
and $E(b,t),$ but was clearly deficient near the zeros of these
quantities, where it led to a flattening or step in $j$ near its zeros.
\ej
\c{\bf V. RESULTS AND DISCUSSION}

In presenting results in Figures 2 through 13, all quantities are given
in natural units, so that
spatial coordinates are measured in $1/m$ and momentum coordinates in
$m.$  The electromagnetic four-potential is presented as
$a(b,t)=eA(b,t)/m^3,$ the electric field as $e(b,t)=eE(b,t)/m^4,$ and the
current as $j(b,t) \to ej(b,t)/m^5.$   The cases chosen for presentation
here parallel parameter sets picked in the study of the infinite-volume
situation [10].  We shall generally show the current and
the electric field averaged over the cylinder cross section according to
$$Q(t) \equiv {1 \over \pi R^2} \int_0^\infty d\b\ q(\b,t),\eqno(27)$$
where $q(\b,t)=j(\b,t)$ or $e(\b,t).$  This, as we shall see, converges
much faster as the number of Fourier-Bessel components $n_{max}$ is
increased than do, say, the on-axis values.

Figures 2 and 3 present the transversely-integrated current and
electric field for unrenormalized charge $e^2 = 10,$ on-axis initial
electric field $e(b=0,\, t=0) = 6,$ and cylinder radii $R=2$ and 5.  The
large value of the charge chosen here has the advantage, given our need
for very lengthy computer time, that back-reaction effects have a chance
to make themselves felt within the relatively short time interval
$t < 4$ (see Ref.  [10]).  As will be generally true below, the electric
field integrated over the cylinder cross section is very similar for the
field-theory approach and the transport calculation.  Of course, this is
partly the case because, for the small radii considered here, the
electric field oscillations are largely driven by cylinder radius, not by
back-reaction.  Thus, by $R = 10$ this agreement begins to break down.
On the other hand, $R = 10$ stretches the range of reliability of our
calculation, since we must require that the ``transverse momentum''
$n_{max}\pi/R$ be on the order of 10 to 20 in order to cover the relevant
momentum distributions and to yield a valid regularization scheme using
Eq. (18).  Since in practice we are limited to $n_{max} \le 20,$ this
implies a breakdown in our procedures for $R$ much beyond 5 or so.  The
transversely-integrated currents as calculated from field theory
are substantially more oscillatory than their transport-theory
counterparts, which is not surprising in view of the infinite-volume
results [10].  The initial sharp rise in the current occurs in both
cases, though with rather different values; thereafter there is
reasonable similarity between the two methods if one takes into account
an effective time-averaging over the field-theory fluctuations.  It is
worth noting [19] that the initial rapid variations in the electric field
in the quantal case may ``shake off'' charges by means of the fast
fluctuations, an effect not included in the transport approach here,
where only pair tunneling is included.

In an attempt to make some connection
with the infinite-volume results, we show $j$ and $e$ on the cylinder
axis for $e^2 = 10,$
initial field $e(b=0,\, t=0) = 6,$ and $R=2$ and 5 in Figures 4 and 5.
Agreement here between field theory and transport formalism
is much less satisfactory than the situation for the
integrated values:  The electric field on-axis remains
similar for the two cases, again because its values are essentially
determined by the initial condition and the cylinder radius, but the
current differs by nearly two orders of magnitude.
This is almost certainly in large part a numerical breakdown brought
about by the difficulty in
using a large enough number of Fourier-Bessel components to achieve
reliable results for the on-axis case. (For Figures 2 through 5 we used
$n_{max} = 20.$)  Since $J_0(0)=1,$ all the
components add coherently to produce the $b=0$ values, and thus these
are very sensitive to limitations in $n_{max}.$  On the other
hand, the on-axis value of the current is
weighted with a factor of $b$ and hence does not really contribute when
global features are considered.

The corresponding picture for a smaller charge, namely, $e^2 = 4,$ with
initial field $e(b=0,\, t=0) = 7,$ and again for $R = 2$ and 5 is shown
in Figures 6 through 9 (where we have worked with $n_{max} = 10$ in
order to allow for the calculation to extend out to $t = 30,$ and
encompass several cycle times).  As is known from the infinite-volume
case [10], the field-theory and transport-formalism results eventually
drift out of phase, a condition that would be improved by incorporating
Bose enhancement into the transport equation [10].  Again, if one were to
average over the fluctuations present in each cycle of the current,
agreement between the two methods is quite reasonable for the integrated
values, and not very good for on-axis results.

In Figures 10 and 11 we compare, for $e^2 = 10,\ e(b=0,\, t=0) = 6,$
and $R = 2$ and 5, results for two different
numbers of Fourier-Bessel components, $n_{max} = 10,$ $dt = 2 \times
10^{-4},$ 401 $k$-grid points, and $\Lambda = 20$ (dashed curve) vs.
$n_{max} = 20,$ $dt = 10^{-4},$ 1,001 $k$-grid points, and
$\Lambda = 40$ (solid curve) for the transversely-integrated current and
electric field.  The same comparisons for the on-axis quantities are shown
in Figures 12 and 13.  Since we have here a doubling in this upper
limit for the number of transverse modes $n_{max},$
the reasonable agreement for the integrated values suggests
that the results there for $n_{max} = 20$ are quantitatively reliable.
Obviously this is not yet the case for the on-axis quantities---notably
the current, which differs by two orders of magnitude for $t > 2$---which
at best have qualitative validity.  It is perhaps not surprising that the
less numerically reliable field-theory result (dashed curve)
more closely resembles the transport calculation,
since its inferior numerics automatically perform some sort of average
over the quantal fluctuations of the field-theory result.  Such an
averaging would otherwise only properly be performed by doing the
field-theory computation with much more refined parameters than are
presently accessible and then averaging these results over time
systematically, or---better still---averaging over momentum distributions
$f_{k,nn'}(t)$ before going on to compute final physical values [10].

In sum, the present results indicate that back-reaction in a finite
volume allows for the replacement of the full field-theory calculation
by the much simpler transport formalism at least at a loose, qualitative
level for quantities averaged over the transverse direction.
It is perhaps not
surprising that the very close linkage between field theory and transport
formalism found in the case of infinite volume [6-10] is lost here:  The
values of the radius $R$ that we deal with are comparable with the particle
mass, so that quantal effects should be expected to enter, and the
spatial variability of the field is such that the ``local'' Schwinger
form of Eq. (23) is not likely to prove adequate.  This is in contrast to
the infinite-volume situation [6-10], where the temporal variability of
the physical quantities is measured by cycle times roughly an order of
magnitude larger than $R$ here.

This rough, qualitative correspondence between field theory and transport
formalism may also apply for the physical quantities in greater detail as
functions of the transverse
parameter $b,$ but present computer capabilities do not yet allow one to
establish this.  In all likelihood the correspondence will be
strengthened somewhat by the consideration of the entire physical context,
including particle interaction, irregular geometry, and the like.  Thus
for many practical purposes one may be able to replace a field-theory
approach by a transport one at the level of qualitative behavior, and
use the latter for further study of pre-equilibrium parton production.
\vfill\eject
It is a pleasure to acknowledge many valuable exchanges on the subject
matter of this paper with Fred Cooper, Yuval Kluger, Emil Mottola, and
Ben Svetitsky.  At the start of the study I also had useful conversations
on it with Gideon Dror and Stefan Graf.  There were also helpful
exchanges on this work at the Workshop on Parton Production and Transport
in the Quark-Gluon Plasma held under the auspices of the European Centre
for Theoretical Studies in Nuclear Physics and Related Areas (ECT*) in
Trento, Italy, in October 1994.
This work was carried out with partial support from the
U.S.-Israel Binational Science Foundation and from the Ne'eman Chair in
Theoretical Nuclear Physics at Tel Aviv University.

\ej
 {\bf References:-}
\v
\baselineskip 12pt
\parskip 0pc
\parindent 1pc
\hangindent 2pc
\hangafter 10

\item{1.}  A. Bia\l as and W. Czy$\dot{\rm z},$ Phys. Rev. D {\bf 30},
2371 (1984); {\it ibid.} {\bf 31}, 198 (1985); Z. Phys. {\bf C28}, 255
(1985); Nucl. Phys. {\bf B267}, 242 (1985); Acta Phys. Pol. {\bf B17},
635 (1986).
\v
\item{2.}  G. Gatoff, A.K. Kerman, and T. Matsui, Phys. Rev. D {\bf
36}, 114 (1987).
\v
\item{3.}  S.A. Fulling, {\sl Aspects of Quantum Field Theory in Curved
Space-Time} (Cambridge University Press, Cambridge, 1989).
\v
\item{4.}  N.D. Birrell and P.C.W. Davies, {\sl Quantum Fields in Curved
Space} (Cambridge University Press, Cambridge, 1982).
\v
\item{5.}  Ya.B. Zel'dovich, in {\sl Magic Without Magic: John Archibald
Wheeler,} ed. J. Klauder (Freeman, San Francisco, 1972) p. 277.
\v
\item{6.}  Y. Kluger, J.M. Eisenberg, B. Svetitsky, F. Cooper, and E.
Mottola, Phys. Rev. Lett. {\bf 67}, 2427 (1991).
\v
\item{7.}  Y. Kluger, J.M. Eisenberg, B. Svetitsky, F. Cooper, and E.
Mottola, Phys. Rev. D {\bf 45}, 4659 (1992).
\v
\item{8.}  F. Cooper, J.M. Eisenberg, Y. Kluger, E. Mottola, and B.
Svetitsky, Phys. Rev. D {\bf 48}, 190 (1993).
\v
\item{9.}  J.M. Eisenberg, Y. Kluger, and B. Svetitsky, Acta Phys. Pol.
 {\bf B23}, 577 (1992).
\v
\item{10.}  Y. Kluger, J.M. Eisenberg, and B. Svetitsky, Int. J. Mod.
Phys. E {\bf 2}, 333 (1993).
\v
\item{11.}  J. Schwinger, Phys. Rev. {\bf 82}, 664 (1951).
\v
\item{12.}  C. Best and J.M. Eisenberg, Phys. Rev. D {\bf 47}, 4639
(1993).
\v
\item{13.}  J. Rau, Phys. Rev. D, in press.
\v
\item{14.}  F. Cooper and Y. Kluger, private communication.
\v
\item{15.}  F. Cooper, S. Habib, Y. Kluger, E. Mottola, J.P. Paz, and
P.R. Anderson, Phys. Rev. D {\bf 50}, 2848 (1994).
\v
\item{16.} J. Ambj\o rn and S. Wolfram, Ann. Phys. {\bf 147}, 33
(1983).
\v
\item{17.}  F. Cooper and E. Mottola, Phys. Rev. D {\bf 40}, 456
(1989).
\v
\item{18.}  W.H. Press, B.P. Flannery, S.A. Teukolsky, and W.T.
Vetterling, {\sl Numerical Recipes} (Cambridge University Press,
Cambridge, 1982).
\v
\item{19.}  Y. Kluger, private communication.
\ej
\baselineskip 15 pt
\parskip 1 pc \parindent 0 pc
\noindent {\bf Figure captions:-}

FIG 1.  Cylindrical geometry considered here.  On the left side of the
cylinder are shown typical actions of forces on negative and positive
charges.  Since charges of opposite sign are evenly distributed and
their transverse motion is the same, they neutralize each other insofar
as transverse contributions to the current are concerned.

FIG 2.  Comparison between field-theory results (left-hand side) and
transport theory calculation (right-hand side) for quantities integrated
over the cylindrical cross section [i.e., tranversely---see Eq. (27)].
Here the charge is $e^2=10,$ the initial electric field on the cylinder
axis is $E(b=0,\,t=0)=6,$ and the cylinder radius is $R=2,$ all in
natural units (see Section IV).

FIG 3.  Same as Figure 2, but for a cylinder radius $R=5.$

FIG 4.  Comparison between field-theory results (left-hand side) and
transport theory calculation (right-hand side) for quantities on the
cylinder axis.  The charge is $e^2=10,$ the initial electric field on
axis is $E(b=0,\,t=0)=6,$ and the cylinder radius is $R=2$ in
natural units (see Section IV).

FIG 5.  Same as Figure 4, but for a cylinder radius $R=5.$

FIG 6.  Comparison between field-theory results (left-hand side) and
transport theory calculation (right-hand side) for quantities integrated
over the cylindrical cross section.
Here the charge is $e^2=4,$ the initial electric field on the cylinder
axis is $E(b=0,\,t=0)=7,$ and the cylinder radius is $R=2$ in natural units.

FIG 7.  Same as Figure 6, but for a cylinder radius $R=5.$

FIG 8.  Comparison between field-theory results (left-hand side) and
transport theory calculation (right-hand side) for quantities on the
cylinder axis.  The charge is $e^2=4,$ the initial electric field on
axis is $E(b=0,\,t=0)=7,$ and the cylinder radius is $R=2$ in
natural units (see Section IV).

FIG 9.  Same as Figure 8, but for a cylinder radius $R=5.$

FIG 10.  Comparison between computation for integrated $J(t),\ A(t),\
E(t)$ [see Eq. (27)] using time interval
$dt=10^{-4},$ 1,001 longitudinal grid points, 20 Fourier-Bessel
components ($n_{max}=20$), and cutoff $\Lambda=40$ taking several weeks
(solid line) with a less accurate version having $dt=2 \times 10^{-4},$
401 grid points, $n_{max}=10,$ and $\Lambda=20,$ taking only a few days.
The physical parameters here are the same as for Figure 2.

FIG 11.  Same comparison as for Figure 10, but with cylinder radius
$R=5.$

FIG 12.  Same comparison as in Figure 10, but for on-axis quantities
$j(b=0,\, t),\ a(b=0,\, t),\ e(b=0,\, t).$

FIG 13.  Same comparison as for Figure 12, but with cylinder radius $R =
5.$

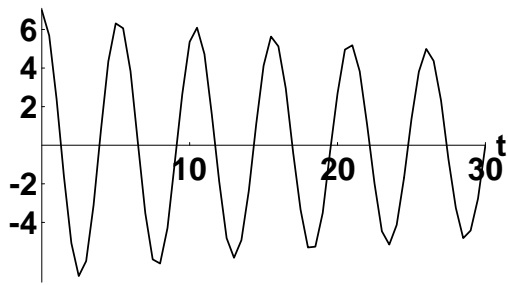
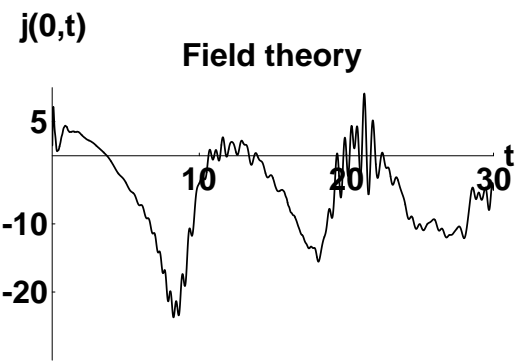
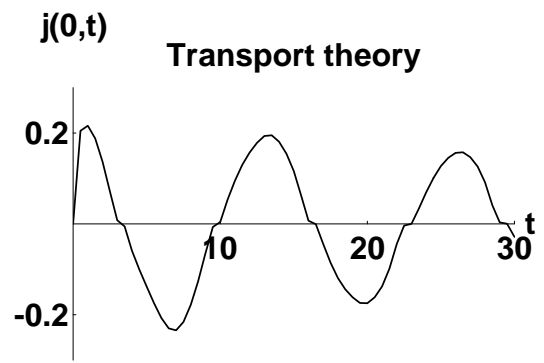
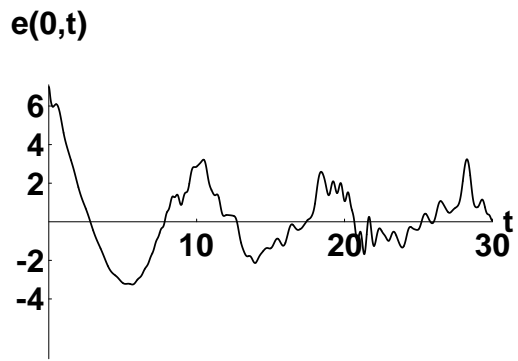
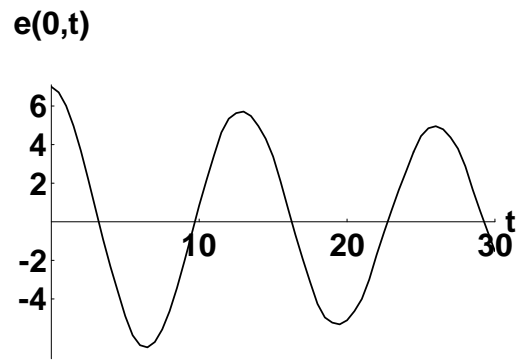
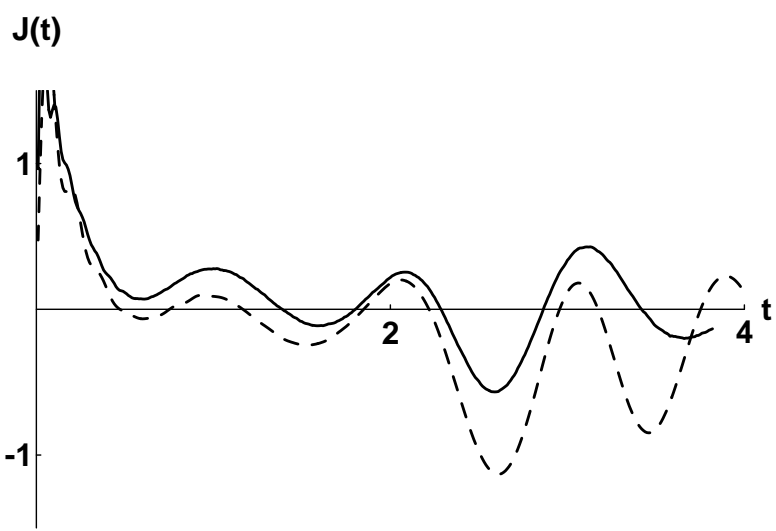
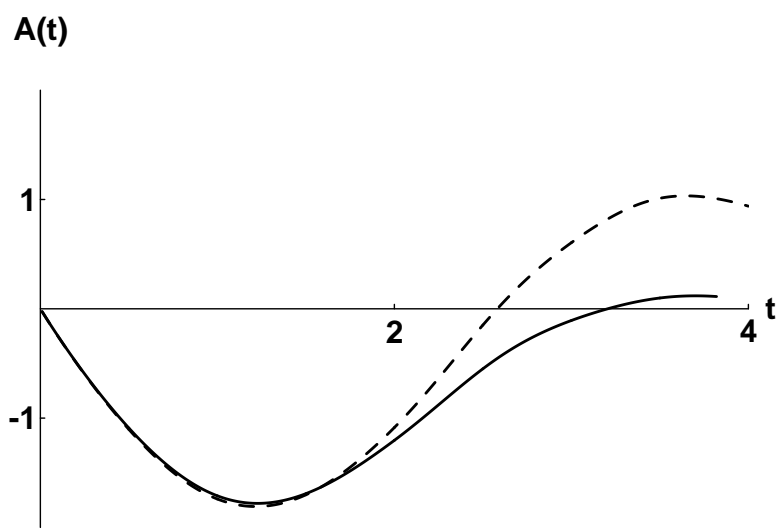
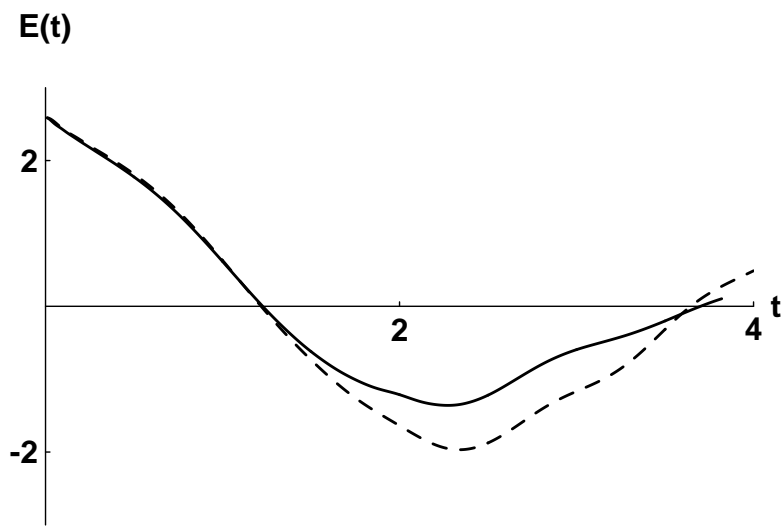
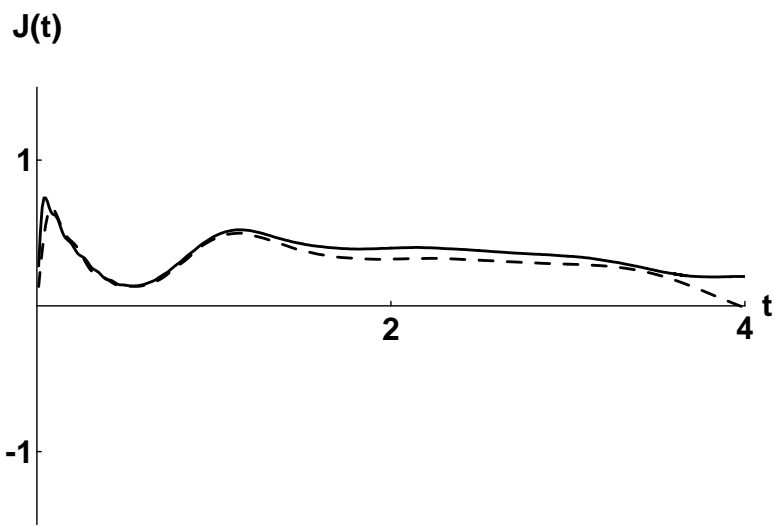
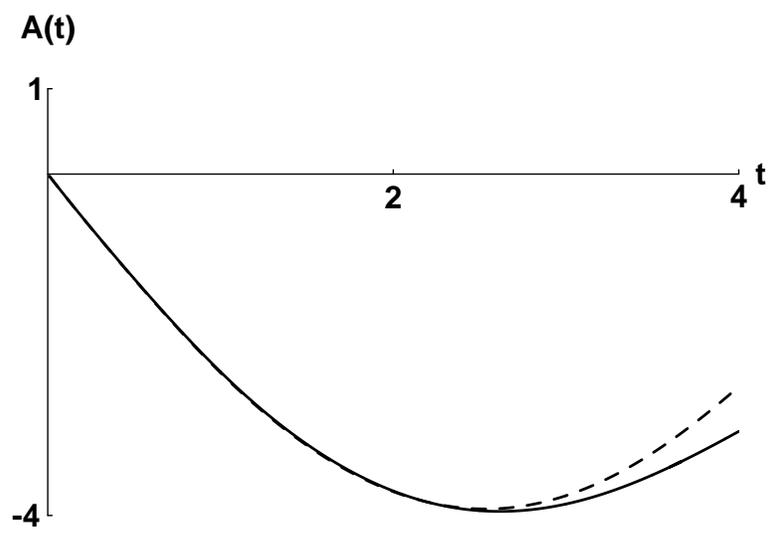
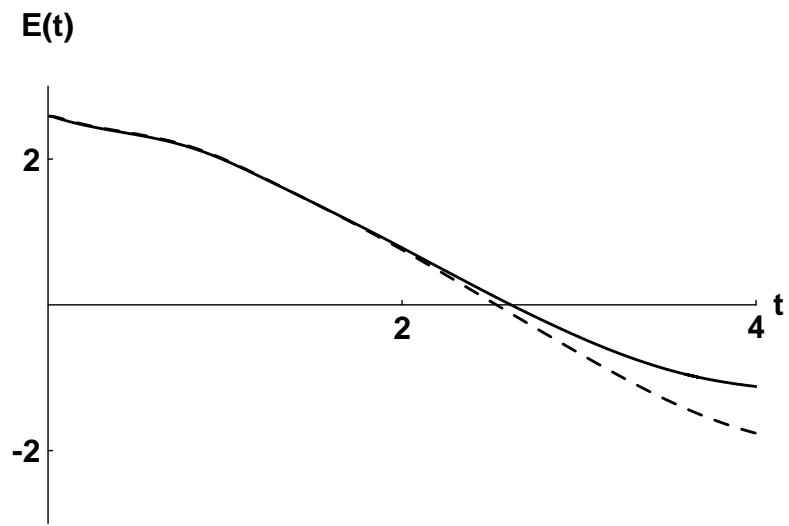
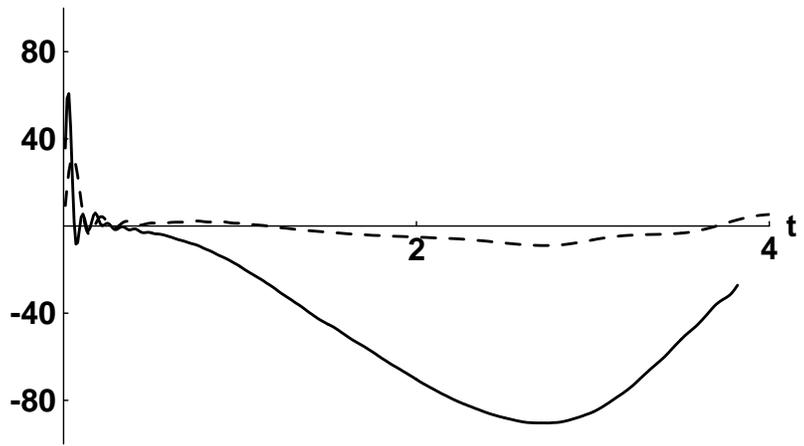
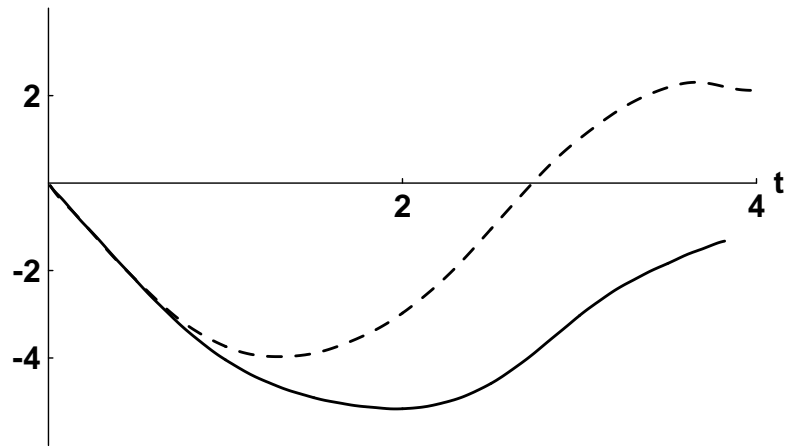
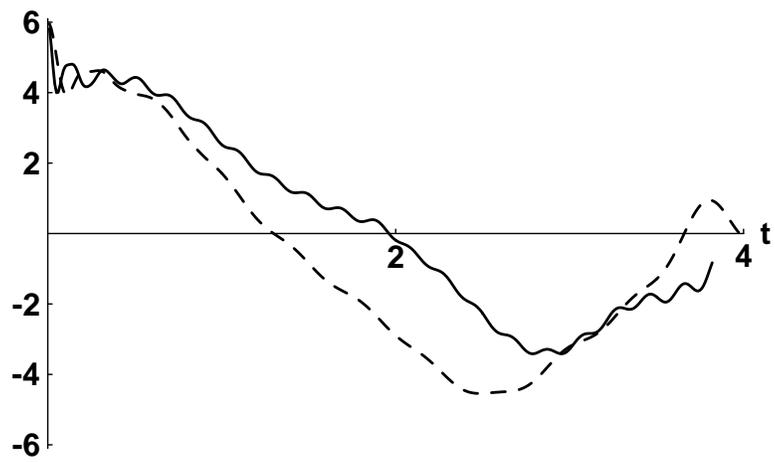
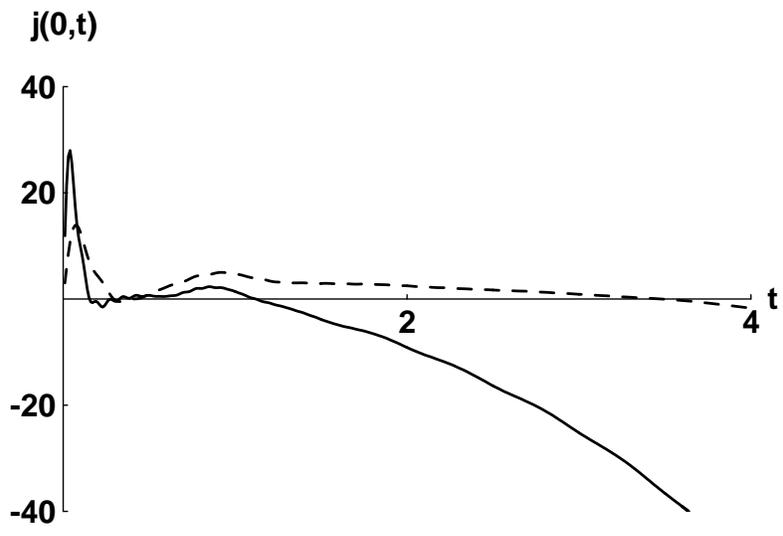
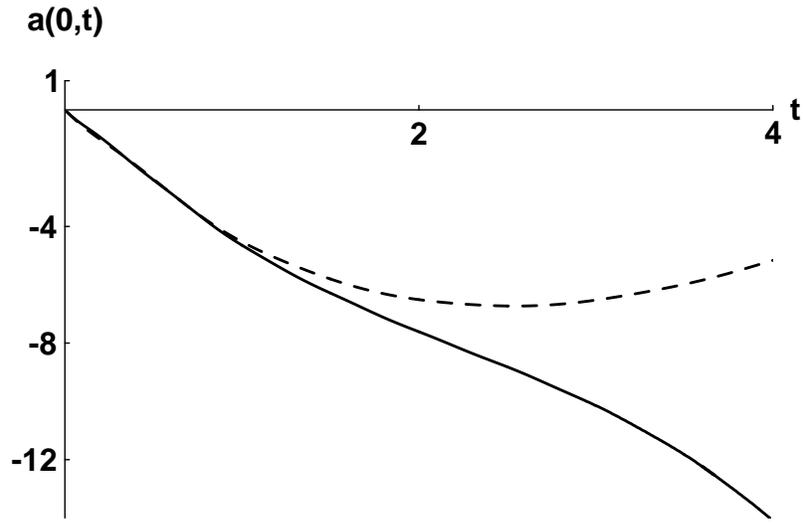
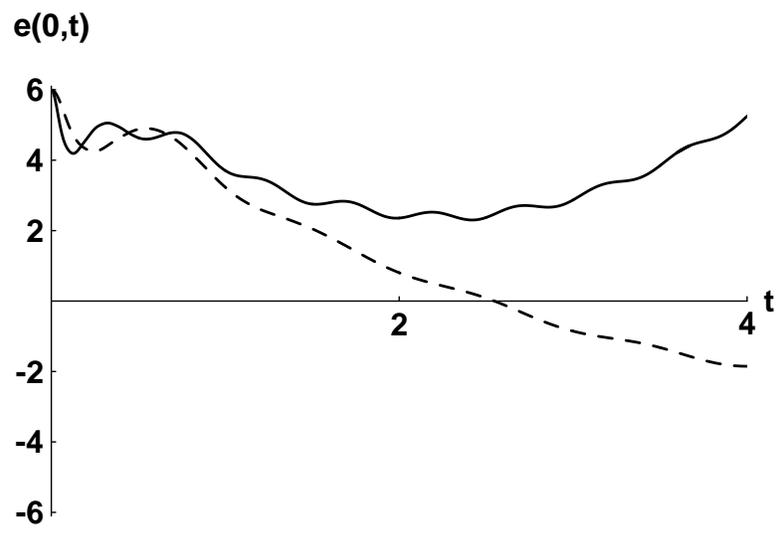
\bye